\documentclass[twocolumn,showpacs,preprintnumbers,amsmath,amssymb]{revtex4}

\usepackage{citesort}
\usepackage{graphicx}
\usepackage{dcolumn}
\usepackage{bm}

\begin{document}
\title{Predictions for the cusp in $\eta\to3\pi^{0}$ decay}
\preprint{HISKP--TH--08/22}

\author{C.-O.~Gullstr\"om$^1$, A.~Kup\'s\'c$^1$\footnote{e-mail: Andrzej.Kupsc@physics.uu.se} and 
A.~Rusetsky$^2$\footnote{e-mail: rusetsky@itkp.uni-bonn.de}}
\affiliation{ 
$^1${Department  of  Physics  and  Astronomy,  Uppsala
    University, Box 516, 75120 Uppsala, Sweden} 
\\
$^2${Helmholtz-Institut  f\"ur Strahlen- und Kernphysik and
Bethe Center for Theoretical Physics, 
    Universit\"at     Bonn,     D-53115    Bonn,
    Germany}
}

\date{\today}
            
\begin{abstract}

A    realistic    estimate    of    the    cusp    effect    in    the
$\eta\rightarrow3\pi^{0}$ decay  is required for  the forthcoming high
precision experiments. The predictions for the size of this effect are
given within the framework of nonrelativistic effective field theory.
\end{abstract}


\maketitle

\section{Introduction}

The  physical region in  $M_{\pi^0\pi^0}$ invariant  mass distribution
for  $\eta\to  3\pi^0$  decay   extends  below  the  charged  two-pion
threshold.  It means  that a cusp structure should  be visible in this
distribution  around $2M_{\pi^\pm}$,  in analogy  with  the pronounced
cusp  in $K^+\to\pi^+\pi^0\pi^0$  decay, observed  recently  by NA48/2
collaboration~\cite{Batley:2005ax}. In  this paper, in  particular, it
has been shown  that measuring charged kaon decays  in the cusp region
enables  one  to  precisely  determine  $S$-wave  $\pi\pi$  scattering
lengths   $a_0$   and   $a_2$,   provided  an   accurate   theoretical
parameterization of the invariant  mass distribution in terms of these
scattering                          lengths                         is
known~\cite{Cabibbo:2004gq,cabibboisidori,prades,cuspwe} (The strong 
impact of the unitarity cusp on $\pi^0\pi^0$ scattering was already 
mentioned in Ref.~\cite{MMS}.).     Moreover,
the same  logic applies to the  neutral kaon decays  into three pions,
which  have been  studied in  the recent  experiment~\cite{kTeV}.  The
theoretical  framework for  analysis  of the  neutral  kaon decays  is
provided   in   Refs.~\cite{cabibboisidori,prades,cuspwe0}   and   the
systematic inclusion  of the  electromagnetic effects both  in charged
and  neutral kaon  decays is  considered in  Ref.~\cite{cuspwerad}. We
further mention  that the  general structure of  the amplitude  in the
neutral  kaon  decays  is   similar  to  the  $\eta\to  3\pi^0$  decay
amplitude. For  this reason, e.g., the two-loop  representation of the
amplitude   in  terms  of   the  $\pi\pi$   effective-range  expansion
parameters, 
derived  in Ref.~\cite{cuspwe0},  can be directly  used to
predict the  cusp in the $\eta\to  3\pi^0$ decay, which  is studied in
KLOE,     Crystal      Ball   and WASA   collaboration     experiments
\cite{Tippens:2001fm,Ambrosino:2007wi,Bashkanov:2007iy}.

It should be pointed out that the two-loop formula for the
kaon decay amplitudes, which was mentioned above, have been obtained
in Refs.~\cite{cuspwe,cuspwe0} within the non-relativistic effective
field theory framework. This framework  is ideally suited for 
parameterizing the final-state interactions in terms of the 
{\em $\pi\pi$ scattering lengths} (effective-range parameters, in general),
whereas the expansion of the amplitudes in Chiral Perturbation Theory (ChPT)
is performed in powers of the {\em quark masses} and is less convenient for
expressing the amplitude in the cusp region in terms of the observable
quantities. 
(Note that, aside from the 
three-pion decays of charged and neutral kaons, the non-relativistic approach 
has been successfully applied recently to study of the $K_{e4}$ 
decays~\cite{ke4}.)

The aim of the present paper is to use the two-loop representation,
derived in Ref.~\cite{cuspwe0}, to estimate the size of the cusp 
in the invariant mass distribution for $\eta\to 3\pi^0$ decays. 
This will finally allow one to judge, whether the forthcoming high-precision
experiment will be able to see the cusp structure in the amplitude.
Note that the cusp effect in the $\eta\to 3\pi^0$ decay has been addressed 
already in various settings, e.g. in Refs.~\cite{BelinaDipl,NisslerPhD,Ditsche}.

In addition, we shall apply the same framework to study the experimental
extraction of the slope parameter
for the decay into three neutral pions. At present, the theory and experiment
have not yet converged to a common denominator for this parameter. 
ChPT at one
loop in the isospin symmetry limit~\cite{Gasser:1984pr}
predicts a different sign
for this parameter as compared to the experimentally measured one.
At two loops, the sign of this quantity is no more
 fixed due to the large error bars coming
from the unknown low-energy constants in ChPT~\cite{Bijnens:2007pr} albeit the
central value is still positive
(The isospin-breaking corrections at one loop have been calculated 
in Refs.~\cite{Wyler,Deandrea,Ditsche} and are found to be small.).
However, the predicted sign in Ref.~\cite{Borasoy:2005du}
where the calculations were done in 
the framework of unitarized ChPT, as well as the sign emerging in
dispersive calculations~\cite{Kambor,GasserBijnens},  
agree with the existing
experimental data. We believe that in the forthcoming high-precision 
measurements of the slope parameter 
it will be very important to use as accurate a 
parameterization of the
decay amplitude, as possible. The parameterization should be 
based on solid theoretical ground and, in particular, should 
take into account the cusp phenomenon which emerges at the physical values
of the pion masses.

\section{Theoretical framework}

Below we mainly follow the notations from  Ref.~\cite{cuspwe0}.  The tree-level
amplitudes are expressed in terms of the kinetic energies $X_i$
\begin{equation}
X_i=E_i-M_{\pi^0}\, ,
\end{equation}
where $E_i$ denote the pion energies in the eta rest frame. 
Up to the quadratic terms,
\begin{eqnarray}\label{eq:tree}
  {\cal M}_{000}^{\rm tree}&=&K_0+K_1(X_1^2+X_2^2+X_3^2)\, ,\nonumber\\[2mm]
  {\cal M}_{+-0}^{\rm tree}&=&L_0+L_1X_3+L_2X_3^2+L_3(X_1-X_2)^2\, ,
\end{eqnarray}
where $L_i,K_i$ are the effective couplings in the non-relativistic Lagrangian
that describe $\eta\to 3\pi$ decays at tree level. Note that we use the same 
notation for these couplings as in Ref.~\cite{cuspwe0}, where they denote the
couplings describing the 3-pion decays of the neutral kaons.

Assuming  $\Delta I=1$ rule in the $\eta\to 3\pi$ vertex, the isospin
symmetry relates the amplitudes for $\eta\to 3\pi^0$ and $\eta\to \pi^+\pi^-\pi^0$ 
(we use Condon-Shortley phase convention)
\begin{eqnarray}
\label{eq:constraints}
&&{\cal M}_{000}(s_1,s_2,s_3)=-{\cal M}_{+-0}(s_1,s_2,s_3)
\nonumber\\[2mm]
&-&{\cal M}_{+-0}(s_2,s_3,s_1)
-{\cal M}_{+-0}(s_3,s_1,s_2)\, .
\end{eqnarray}
At tree level, this allows one to express the couplings $K_i$ through $L_i$
\begin{eqnarray}
\label{eq:constraints-tree}
K_0&=&-(3L_0+L_1Q-L_3Q^2)\, ,\nonumber\\[2mm]
K_1&=&-(L_2+3L_3)\, ,
\end{eqnarray}
where $Q=M_\eta-3M_{\pi^0}$. 

In general, $\eta\to 3\pi$ decay  amplitudes are given in a form of a sum
of the tree, one-loop, two-loop,~$\ldots$ contributions
${\cal M}_{000}={\cal M}_0^{\rm tree}+{\cal M}_0^{\rm 1-loop}
+{\cal M}_0^{\rm 2-loops}+\cdots$, and similarly for ${\cal M}_{+-0}$.
 The pertinent (rather lengthy) 
expressions are given in Ref.~\cite{cuspwe0}. We do not display them here.
It can be checked that these amplitudes in the isospin symmetry limit 
explicitly obey the 
constraints~(\ref{eq:constraints}) at one- and two-loop level.

We wish to stress that the representations given in 
Refs.~\cite{cuspwe,cuspwe0} should be understood as a parameterization 
which should be fit to the data. In other words, the constants 
$L_i,a_0,a_2,\ldots$
are considered as free parameters to be fixed from the 
fit. In this paper, we however make an attempt to {\em predict} the size of the
cusp -- fitting first the tree-level amplitude in order to determine $L_i$
and then using one- and two-loop representation to produce the cusp in the 
synthetic data. In doing this, we have fixed $a_0,a_2$ to their 
theoretical values~\cite{CGL} and neglected isospin breaking in the 
derivative 4-pion couplings, as well as the shape parameter and the
$P$-waves.

The matching of $L_i$ is done to:
\begin{itemize}
\item [1)] 
The tree-level amplitude $\eta\to\pi^+\pi^-\pi^0$ in ChPT. 
According to Eq.~(\ref{eq:constraints-tree}),
the overall normalization of the amplitude does not play a role, only the
slopes matter. The result is given by
\begin{eqnarray}
L_0&=&(4M_{\pi^0}^2-3(M_\eta-M_{\pi^0}))^2/(M_\eta^2-M_{\pi^0}^2)\, ,\nonumber\\[2mm]
L_1&=&6M_\eta/(M_\eta^2-M_{\pi^0}^2)\, ,\nonumber\\[2mm]
L_2&=&L_3=0\, .
\end{eqnarray}

\item[2)]
The experimental amplitude extracted by KLOE   
collaboration~\cite{Ambrosino:2008ht}.
\end{itemize}

In order to carry out the matching to the KLOE data, it is useful to introduce
Dalitz variables for $\eta\to\pi^+\pi^-\pi^0$ decay
\begin{equation}
x=\sqrt{3}(X_1-X_2)/Q\, ,\quad
y=3X_3/Q-1\, .
\end{equation}
For the decay $\eta\to 3\pi^0$ one defines the variable
\begin{equation}
z= x^2+y^2\, .
\label{eqn:z}
\end{equation}
The phenomenological parameterization of the amplitude
is given by
\begin{eqnarray}\label{eq:KLOE}
{\cal M}_{+-0}&=&A_c(1+\alpha y+\beta y^2+ \gamma x^2)\, ,
\end{eqnarray}
with $\alpha,\beta,\gamma$ being complex quantities. The matching
of Eq.~(\ref{eq:tree}) to the {\em real part} of Eq.~(\ref{eq:KLOE}) 
yields
\begin{eqnarray}\label{eq:KLOEmatching}
L_0&=&A_c(1-\mbox{Re}\,\alpha+\mbox{Re}\,\beta)\, ,\nonumber\\[2mm]
L_1&=&3A_c(\mbox{Re}\,\alpha-2\mbox{Re}\,\beta)/Q\, ,\nonumber\\[2mm]
L_2&=&9A_c\,\mbox{Re}\,\beta/Q^2\, ,\nonumber\\[2mm]
L_3&=&3A_c\,\mbox{Re}\,\gamma/Q^2\, .
\end{eqnarray}
The right-hand side in Eq.~(\ref{eq:KLOEmatching}) is fixed by using
Eq.~(6.4) and table 1 of Ref.~\cite{Ambrosino:2008ht}.
Isospin-breaking corrections in Eqs.~(\ref{eq:KLOE}) and 
(\ref{eq:KLOEmatching}) are consistently neglected.

We would like to mention that the systematic way of fixing the parameters 
of the effective non-relativistic Lagrangian consists in performing a 
{\em simultaneous} fit of the non-relativistic representation
to both charged and neutral invariant mass distributions. The results, which 
are contained in the present paper, should be considered only 
as a rough theoretical estimate of the expected size of the cusp effect in the 
$\eta\to 3\pi^0$ decay.

\begin{figure}
\includegraphics[width=\columnwidth]{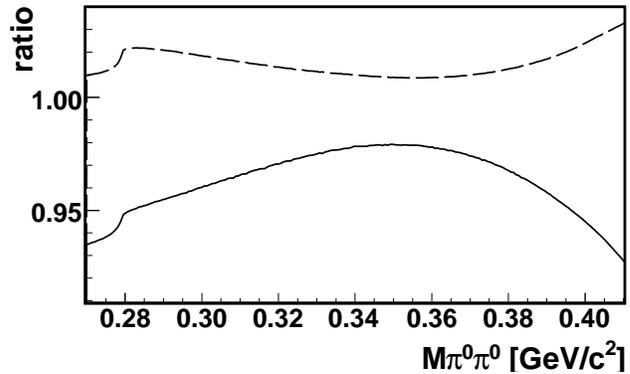}
\caption[]{Invariant mass distribution 
$d\Gamma/dM_{\pi^0\pi^0}$ divided by the phase space, calculated at two loops:
1) Matching to ChPT at tree level (dashed line);
2) Matching to the KLOE parameterization~\cite{Ambrosino:2008ht} (solid line).
 }
\label{fig:1}
\end{figure}

\begin{figure}
\includegraphics[width=\columnwidth]{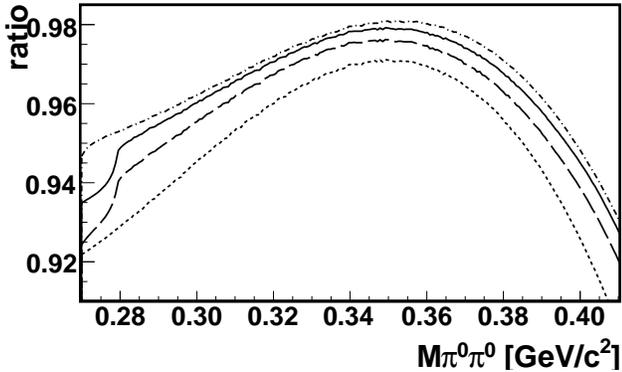}
\caption[]{Invariant mass distribution 
$d\Gamma/dM_{\pi^0\pi^0}$ divided by the phase space,
with the couplings matched to the KLOE 
parameterization~\cite{Ambrosino:2008ht}: 
1) Tree level (dotted line);
2) One loop (dashed line);
3) Two loops (solid line);
4) Two loops, assuming $M_{\pi^\pm}=M_{\pi^0}$ (dot-dashed line).
}
\label{fig:2}
\end{figure}

\begin{figure}
\includegraphics[width=\columnwidth]{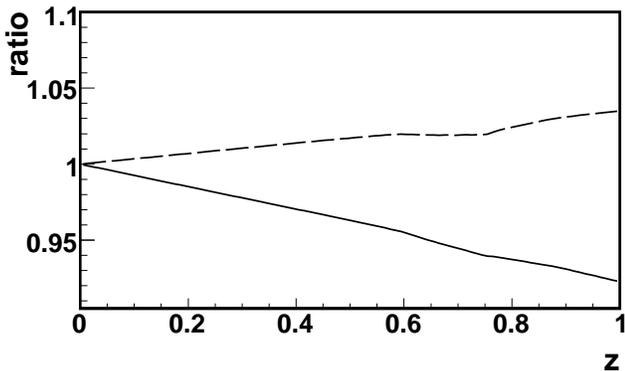}
\caption[]{Differential decay rate $d\Gamma/dz$ divided by the phase space
 at two loop:
1) Matching to ChPT at tree level (dashed line);
2) Matching to the KLOE parameterization~\cite{Ambrosino:2008ht}.
}
\label{fig:3}
\end{figure}

\begin{figure}
\includegraphics[width=\columnwidth]{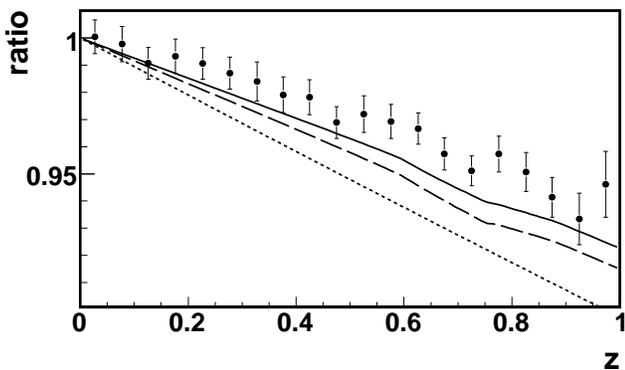}
\caption[]{  Calculated $d\Gamma/dz$ divided  by the
  phase    space    and   comparison    with    Crystal   Ball    data
  \cite{Tippens:2001fm}.     Couplings    matched    to    the    KLOE
  parameterization~\cite{Ambrosino:2008ht}.   1)  Tree  level  (dotted
  line); 2) One loop (dashed line); 3) Two loops (solid line).  }
\label{fig:4}
\end{figure}

\section{Results}

In Fig.~\ref{fig:1} we display the calculated invariant mass distribution
for $\eta\to 3\pi^0$ decay, divided by the phase space. The decay amplitude is
normalized in the center of the Dalitz plot
\begin{equation}
|{\cal M}_{000}(s_0,s_0,s_0)|^2=1\, ,\quad s_0=\frac{M_\eta^2}{3}+M_{\pi^0}^2\, .
\end{equation}
We display the result for $L_i,K_i$ matched to the tree-level result of ChPT,
or to the KLOE amplitude~\cite{Ambrosino:2008ht}. The resulting cusp
in both cases amounts roughly up to a $2\%$ effect.
We would like to mention that the sign of the cusp effect is fixed by
the isospin symmetry, see Eqs.~(\ref{eq:constraints}) and 
(\ref{eq:constraints-tree}) and is thus a robust theoretical prediction.

In order to check the convergence of the method, in Fig.~(\ref{fig:2})  
we show the invariant mass distribution calculated at tree level, one and
two loops, with the couplings $L_i$ matched to the KLOE amplitude.
It is seen that the shape of the cusp does not change much from one- to
two-loop calculations, indicating at a rather robust prediction for a size
of this effect.  

Figure~\ref{fig:3} contains our prediction for the differential
decay rate in the variable $z$ -- again with 
$L_i,K_i$ matched either to the tree-level result of ChPT,
or to the KLOE amplitude. As expected, the slope parameter in the former case
has the opposite sign as compared to the experimentally observed. 
Apart from a small dip around $z\simeq 0.75$, corresponding to the cusp,
the differential decay rate is seen to be fairly linear in $z$.

The convergence of the loop expansion for the differential decay rate in the
variable $z$ and the comparison with the Crystal Ball 
data~\cite{Tippens:2001fm} is shown in Fig.~\ref{fig:4}. As seen, the data
are described quite well.

\section{Conclusions}

Using the two-loop parameterization of the $\eta\to 3\pi$ decay 
amplitudes~\cite{cuspwe0}, we have shown that the size of the cusp effect
in the invariant mass distribution for $\eta\to 3\pi^0$ process
amounts up to around $2\%$. Despite such
tiny effect, one may expect that forthcoming high-precision experiments
at Crystal Ball, KLOE and WASA-at-COSY with 
about   $10^7$ events in the Dalitz plot will be able to observe it. It 
is however unlikely that one could determine
$\pi\pi$ scattering lengths at a reasonable accuracy from these experiments.

Moreover, the cusp effect modifies the differential decay rate for the 
$\eta\to 3\pi^0$ decay in the variable $z$, producing a dip around
the value $z\simeq 0.75$. We expect that, in order to carry out an accurate
 analysis of the Dalitz plot distributions,
this effect should be taken into account.

Finally, we wish to mention that for the cusp-like structure, which has been
seen recently by the Crystal Ball collaboration experiment at 
MAMI-C~\cite{Prakhov}, the
{\em sign} of the effect is claimed to be different from the theoretical prediction.
To resolve this contradiction, experimental study of the $\eta\to 3\pi^0$ decay with a 
better statistics would be desirable.

\begin{acknowledgments}
We  thank  J.  Gasser,   B.  Kubis,  U.-G.  Mei{\ss}ner,  S.  Prakhov,
J. Bijnens and R.   Nissler for useful discussions.  Partial financial
support  under  the  EU  Integrated Infrastructure  Initiative  Hadron
Physics  Project  (contract  number  RII3--CT--2004--506078)  and  DFG
(SFB/TR  16,   ``Subnuclear  Structure  of   Matter'')  is  gratefully
acknowledged.   This work was  supported by  EU MRTN--CT--2006--035482
(FLAVIA{\it  net}).  We acknowledge  gratefully the  financial support
(by FZ J\"ulich and TR16)  for the participation to the Hadron Physics
Summer School 2008 in Bad Honnef.
\end{acknowledgments}


\end{document}